\documentclass[aps,prc,twocolumn,10pt,superscriptaddress,floatfix,showpacs]{revtex4-1}

\usepackage{graphicx}  
\usepackage{amssymb}   
\usepackage{hyperref}  

\begin{document}

\title{Extraction of the Neutron Electric Form Factor from Measurements of Inclusive Double Spin Asymmetries}

%
%
\author{V.~Sulkosky}\affiliation{Massachusetts Institute of Technology, Cambridge, MA, 02139, USA}\affiliation{University of Virginia, Charlottesville, VA, 22904, USA}
\author{G.~Jin} \affiliation{University of Virginia, Charlottesville, VA, 22904, USA}
\author{E.~Long} \affiliation{University of New Hampshire, Durham, NH, 03824, USA} 
\author{Y.-W.~Zhang} \affiliation{Rutgers University, New Brunswick, NJ, 08901, USA}
\author{M.~Mihovilovic} \affiliation{Jozef Stefan Institute, Ljubljana 1000, Slovenia}
\author{A.~Kelleher} \affiliation{Massachusetts Institute of Technology, Cambridge, MA, 02139, USA}
\author{B.~Anderson} \affiliation{Kent State University, Kent, OH, 44242, USA}
\author{D.W.~Higinbotham}\email[Corresponding author: ]{doug@jlab.org}\affiliation{Thomas Jefferson National Accelerator Facility, Newport News, VA 23606, USA}
\author{S.~\v{S}irca} \affiliation{Faculty of Mathematics and Physics, University of Ljubljana, Ljubljana, 1000, Slovenia}\affiliation{Jozef Stefan Institute, Ljubljana 1000, Slovenia}
\author{K.~Allada} \affiliation{Thomas Jefferson National Accelerator Facility, Newport News, VA 23606, USA}
\author{J.R.M.~Annand} \affiliation{Glasgow University, Glasgow, G12 8QQ, Scotland, United Kingdom}
\author{T.~Averett} \affiliation{The College of William and Mary, Williamsburg, VA, 23187, USA}
\author{W.~Bertozzi} \affiliation{Massachusetts Institute of Technology, Cambridge, MA, 02139, USA}
\author{W.~Boeglin} \affiliation{Florida International University, Miami, FL, 33181, USA}
\author{P.~Bradshaw} \affiliation{The College of William and Mary, Williamsburg, VA, 23187, USA}
\author{A.~Camsonne} \affiliation{Thomas Jefferson National Accelerator Facility, Newport News, VA 23606, USA}
\author{M.~Canan} \affiliation{Old Dominion University, Norfolk, VA, 23508, USA}
\author{G.D.~Cates} \affiliation{University of Virginia, Charlottesville, VA, 22904, USA}
\author{C.~Chen} \affiliation{Hampton University , Hampton, VA, 23669, USA}
\author{J.-P.~Chen} \affiliation{Thomas Jefferson National Accelerator Facility, Newport News, VA 23606, USA}
\author{E.~Chudakov} \affiliation{Thomas Jefferson National Accelerator Facility, Newport News, VA 23606, USA}
\author{R.~De~Leo} \affiliation{Universite di Bari, Bari, 70121 Italy}
\author{X.~Deng} \affiliation{University of Virginia, Charlottesville, VA, 22904, USA}
\author{A.~Deur} \affiliation{Thomas Jefferson National Accelerator Facility, Newport News, VA 23606, USA}
\author{C.~Dutta} \affiliation{University of Kentucky, Lexington, KY, 40506, USA}
\author{L.~El~Fassi} \affiliation{Rutgers University, New Brunswick, NJ, 08901, USA}
\author{D.~Flay} \affiliation{Temple University, Philadelphia, PA, 19122, USA}
\author{S.~Frullani} \thanks{Deceased} \affiliation{Istituto Nazionale Di Fisica Nucleare, INFN/Sanita, Roma, Italy}
\author{F.~Garibaldi} \affiliation{Istituto Nazionale Di Fisica Nucleare, INFN/Sanita, Roma, Italy}
\author{H.~Gao} \affiliation{Duke University, Durham, NC, 27708, USA}
\author{S.~Gilad} \affiliation{Massachusetts Institute of Technology, Cambridge, MA, 02139, USA}
\author{R.~Gilman} \affiliation{Thomas Jefferson National Accelerator Facility, Newport News, VA 23606, USA}\affiliation{Rutgers University, New Brunswick, NJ, 08901, USA}
\author{O.~Glamazdin} \affiliation{Kharkov Institute of Physics and Technology, Kharkov 61108, Ukraine}
\author{S.~Golge} \affiliation{Old Dominion University, Norfolk, VA, 23508, USA}
\author{J.~Gomez} \affiliation{Thomas Jefferson National Accelerator Facility, Newport News, VA 23606, USA}
\author{J.-O.~Hansen} \affiliation{Thomas Jefferson National Accelerator Facility, Newport News, VA 23606, USA}
\author{T.~Holmstrom} \affiliation{Longwood University, Farmville, VA, 23909, USA}
\author{J.~Huang} \affiliation{Massachusetts Institute of Technology, Cambridge, MA, 02139, USA} \affiliation{Los Alamos National Laboratory, Los Alamos, NM, 87545, USA}
\author{H.~Ibrahim} \affiliation{Cairo University, Cairo, Giza 12613, Egypt}
\author{C.W.~de~Jager} \thanks{Deceased} \affiliation{Thomas Jefferson National Accelerator Facility, Newport News, VA 23606, USA}\affiliation{University of Virginia, Charlottesville, VA, 22904, USA}
\author{E.~Jensen} \affiliation{Christopher Newport University, Newport News, VA, 23606, USA}
\author{X.~Jiang} \affiliation{Los Alamos National Laboratory, Los Alamos, NM, 87545, USA}
\author{M.~Jones} \affiliation{Thomas Jefferson National Accelerator Facility, Newport News, VA 23606, USA}
\author{H.~Kang} \affiliation{Seoul National University, Seoul, Korea}
\author{J.~Katich} \affiliation{The College of William and Mary, Williamsburg, VA, 23187, USA}
\author{H.P.~Khanal} \affiliation{Florida International University, Miami, FL, 33181, USA}
\author{P.~King} \affiliation{Ohio University, Athens, OH, 45701, USA}
\author{W.~Korsch} \affiliation{University of Kentucky, Lexington, KY, 40506, USA}
\author{J.~LeRose} \affiliation{Thomas Jefferson National Accelerator Facility, Newport News, VA 23606, USA}
\author{R.~Lindgren} \affiliation{University of Virginia, Charlottesville, VA, 22904, USA}
\author{H.-J.~Lu} \affiliation{Huangshan University, People's Republic of China}
\author{W.~Luo} \affiliation{Lanzhou University, Lanzhou, Gansu, 730000, People's Republic of China}
\author{P.~Markowitz} \affiliation{Florida International University, Miami, FL, 33181, USA}
\author{D.~Meekins} \affiliation{Thomas Jefferson National Accelerator Facility, Newport News, VA 23606, USA}
\author{M.~Meziane} \affiliation{The College of William and Mary, Williamsburg, VA, 23187, USA}
\author{R.~Michaels} \affiliation{Thomas Jefferson National Accelerator Facility, Newport News, VA 23606, USA}
\author{B.~Moffit} \affiliation{Thomas Jefferson National Accelerator Facility, Newport News, VA 23606, USA}
\author{P.~Monaghan} \affiliation{Hampton University , Hampton, VA, 23669, USA}
\author{N.~Muangma} \affiliation{Massachusetts Institute of Technology, Cambridge, MA, 02139, USA}
\author{S.~Nanda} \affiliation{Thomas Jefferson National Accelerator Facility, Newport News, VA 23606, USA}
\author{B.E.~Norum} \affiliation{University of Virginia, Charlottesville, VA, 22904, USA}
\author{K.~Pan} \affiliation{Massachusetts Institute of Technology, Cambridge, MA, 02139, USA}
\author{D.~Parno} \affiliation{Carnegie Mellon University, Pittsburgh, PA, 15213, USA}
\author{E.~Piasetzky} \affiliation{Tel Aviv University, Tel Aviv 69978, Israel}
\author{M.~Posik} \affiliation{Temple University, Philadelphia, PA, 19122, USA}
\author{V.~Punjabi} \affiliation{Norfolk State University, Norfolk, VA, 23504, USA}
\author{A.J.R.~Puckett} \affiliation{Massachusetts Institute of Technology, Cambridge, MA, 02139, USA}
\author{X.~Qian} \affiliation{Duke University, Durham, NC, 27708, USA}
\author{Y.~Qiang} \affiliation{Thomas Jefferson National Accelerator Facility, Newport News, VA 23606, USA}
\author{X.~Qui} \affiliation{Lanzhou University, Lanzhou, Gansu, 730000, People's Republic of China}
\author{S.~Riordan} \affiliation{University of Virginia, Charlottesville, VA, 22904, USA} \affiliation{University of Massachusetts, Amherst, MA, 01006, USA}\affiliation{Stony Brook University, Stony Brook, NY 11794, USA}
\author{A.~Saha} \thanks{Deceased} \affiliation{Thomas Jefferson National Accelerator Facility, Newport News, VA 23606, USA} 
\author{B.~Sawatzky} \affiliation{Thomas Jefferson National Accelerator Facility, Newport News, VA 23606, USA}
\author{M.~Shabestari} \affiliation{University of Virginia, Charlottesville, VA, 22904, USA}
\author{A.~Shahinyan} \affiliation{Yerevan Physics Institute, Yerevan, Armenia}
\author{B.~Shoenrock} \affiliation{Northern Michigan University, Marquette, MI, 49855, USA}
\author{J.~St.~John} \affiliation{Longwood University, Farmville, VA, 23909, USA}
\author{R.~Subedi} \affiliation{George Washington University, Washington, D.C., 20052, USA}
\author{W.A.~Tobias} \affiliation{University of Virginia, Charlottesville, VA, 22904, USA}
\author{W.~Tireman} \affiliation{Northern Michigan University, Marquette, MI, 49855, USA}
\author{G.M.~Urciuoli} \affiliation{Istituto Nazionale Di Fisica Nucleare, INFN/Sanita, Roma, Italy}
\author{D.~Wang} \affiliation{University of Virginia, Charlottesville, VA, 22904, USA}
\author{K.~Wang} \affiliation{University of Virginia, Charlottesville, VA, 22904, USA}
\author{Y.~Wang} \affiliation{University of Illinois at Urbana-Champaign, Urbana, IL, 61801, USA}
\author{J.~Watson} \affiliation{Kent State University, Kent, OH, 44242, USA} 
\author{B.~Wojtsekhowski} \affiliation{Thomas Jefferson National Accelerator Facility, Newport News, VA 23606, USA}
\author{Z.~Ye} \affiliation{Hampton University , Hampton, VA, 23669, USA}
\author{X.~Zhan} \affiliation{Massachusetts Institute of Technology, Cambridge, MA, 02139, USA}
\author{Y.~Zhang} \affiliation{Lanzhou University, Lanzhou, Gansu, 730000, People's Republic of China}
\author{X.~Zheng} \affiliation{University of Virginia, Charlottesville, VA, 22904, USA}
\author{B.~Zhao} \affiliation{The College of William and Mary, Williamsburg, VA, 23187, USA}
\author{L.~Zhu} \affiliation{Hampton University , Hampton, VA, 23669, USA}
\collaboration{Jefferson Lab Hall A Collaboration} \noaffiliation
%


%
%
\begin{abstract}
\begin{description}

\item[Background] Measurements of the neutron charge form factor, $G^n_E$, are
challenging due to the fact that the neutron has no net charge.
In addition, measurements of the neutron form factors must use nuclear targets which
require accurately accounting for nuclear effects.   Extracting 
$G^n_E$ with different targets and techniques provides an important test of 
our handling of these effects.

\item[Purpose] The goal of the measurement was to use an inclusive asymmetry measurement 
technique to extract the neutron charge form factor at a four-momentum transfer of 
$1~(\rm{GeV/c})^2$.    This technique has very different systematic uncertainties 
than traditional exclusive measurements and thus serves as an independent check of
whether nuclear effects have been taken into account correctly.

\item[Method] The inclusive quasi-elastic reaction
$^3\overrightarrow{\rm{He}}(\overrightarrow{e},e')$
was measured at Jefferson Lab.
The neutron electric form factor, $G_E^n$, was extracted
at $Q^2 = 0.98~(\rm{GeV/c})^2$ from ratios of electron-polarization
asymmetries measured for two orthogonal target spin orientations.  
This $Q^2$ is high enough that the sensitivity to $G_E^n$ is not overwhelmed 
by the neutron magnetic contribution, and yet low enough that explicit 
neutron detection is not required to suppress pion production.

\item[Results] The neutron electric form factor, $G_E^n$, was determined to be
$0.0414\pm0.0077\;{(stat)}\pm0.0022\;{(syst)}$; providing the first high precision
inclusive extraction of the neutron's charge form factor.

\item[Conclusions] The use of the inclusive quasi-elastic
$^3\overrightarrow{\rm{He}}(\overrightarrow{e},e')$ with a four-momentum transfer
near $1~(\rm{GeV/c})^2$ has been used to provide a unique measurement of $G^n_E$.
This new result provides a systematically independent validation of the exclusive 
extraction technique results and implies that the nuclear corrections are understood.
This is contrary to the proton form factor where asymmetry and cross-section
measurements have been shown to have large systematic differences.

\end{description}
\end{abstract}


\maketitle

\section{Introduction}

Electromagnetic form factors describe the nucleon's
static electromagnetic structure and provide insight into understanding 
nucleons in terms of their fundamental degrees of freedom.  Of the four electromagnetic form factors 
of the proton and neutron ($G_E^p$, $G_M^p$, $G_{E}^{n}$, and $G_M^n$), the measurement 
of $G_E^n$ is particularly challenging due to its small value and the difficulty in 
obtaining a high-density \lq\lq pure'' neutron target.  Extractions of neutron form factors
have relied on measurements on light nuclei, such as the deuteron or $^3$He, where 
the neutron is bound inside the nucleus.  Experimental methods that provide access
to $G_E^n$ include Rosenbluth separations from an unpolarized deuteron
target~\cite{Galster:1971kv,Platchkov:1989ch} and double-polarization
measurements using either a polarized 
target~\cite{Passchier:1999cj, Warren:2003ma,Geis:2008aa,Becker:1999tw,Bermuth:2003qh,Riordan:2010id,Schlimme:2013eoz} or 
an unpolarized target combined with a
polarimeter to measure the polarization transfer to the recoiling
neutron~\cite{Herberg:1999ud,Ostrick:1999xa,Glazier:2004ny,Plaster:2005cx}.
At low four-momentum-transfer-squared, $Q^2$ from 0.1 to 0.2~$(\rm{GeV/c})^2$, 
inclusive quasi-elastic scattering from a polarized $^3$He target was also
tried~\cite{JonesWoodward:1991ih, Thompson:1992ci}.
However, these early measurements yielded statistical uncertainties comparable with 
the extracted quantity; the sources of theoretical uncertainties were not investigated.
In a later measurement, better statistical precision was obtained, and an
extensive analysis of the systematic uncertainties was performed~\cite{Hansen:1994ba}.
In that analysis, the large variation in the asymmetry predictions revealed a
large model uncertainty at low $Q^2$.  The authors of that paper suggested that
the extraction would be likely to succeed at higher $Q^2$.
We report in this paper an extraction of $G_E^n$ at $Q^2 = 0.98~(\rm{GeV/c})^2$
from measurements of the ratios of two asymmetries in the
$^3\overrightarrow{\rm{He}}(\overrightarrow{e},e')$ reaction
where the $^3$He spin vectors aligned parallel and orthogonal
to the electron beam direction.

\section{Methods}

The measurements were performed at Jefferson Lab in experimental Hall A. 
A longitudinally polarized electron beam of 3.606~GeV was scattered from a
gaseous polarized $^3$He target.  The beam current was between 10~$\mu$A and 
15~$\mu$A, and the helicity of the beam was flipped at a frequency of 30~Hz. 
During the experiment, the beam charge asymmetry was minimized by a beam charge 
feedback system \cite{Androic:2011rha} and was controlled to be less than 100 
parts per million (ppm) per 20-30 min time period.  Interruptions of the beam
were found to have negligible effects on the asymmetry.  As a dedicated beam 
polarization measurement in Hall~A was not conducted during the period the data 
were taken, the average beam polarization was determined from measurements 
taken in Hall~B with a M{\o}ller polarimeter to be $(82\pm2.5)$\%~\cite{JinGe_thesis}.

A polarized $^3$He target was used as an effective polarized neutron target.
The target, made of aluminosilicate glass, consisted of a 
pumping chamber and a target cell.  The spherical pumping chamber was
located above the cylindrical target cell and was connected to the target
cell by a transfer tube.  
The $^3$He nuclei were polarized via spin-exchange optical pumping of a Rb-K 
mixture~\cite{Babcock:2003zz}.
The vapor of the alkali mixture was polarized in the pumping chamber, where the
spin exchange with the $^3$He nuclei occurred.
The 40-cm-long target cell contained $^3$He gas at 12~atm, which provided a luminosity 
of 10$^{36}$~cm$^{-2}$s$^{-1}$.  A small amount ($\simeq$ 2\% in number density) of N$_2$ gas 
was added to the target cell to absorb unwanted photons emitted from the Rb de-excitation
process.  With the aid of spectrally narrowed lasers that increase the light absorption 
efficiency~\cite{Singh:2013nja}, a significant improvement in target polarization 
was achieved compared to previous experiments with similar targets.
The polarization of the cell was measured every 6~hours using nuclear magnetic
resonance, calibrated using electron paramagnetic resonance~\cite
{Romalis:1998ik} polarimetry.  An average in-beam target polarization of $(50.2\pm2.5)$\% 
was achieved.  Additionally, a reference cell that could be filled with either $^3$He, 
N$_2$, or H$_2$ gas, was used to determine the dilution factors for the unpolarized material 
in the cell.

The scattered electrons were detected in the Right High Resolution Spectrometer
(RHRS)~\cite{Alcorn:2004sb}.  The RHRS was located at a forward angle of 17$^{\circ}$ 
with respect to the incident beam direction and its central momentum was set to $3.086~\rm{GeV/c}$.
Thus, the momentum transfer to the target $\left(\vec{q}\right)$
by an electron scattered
into the center of the RHRS acceptance was pointing at an angle of 56$^\circ$
with respect to the incident beam direction.
Scattered electrons traveled through the RHRS by passing through a pair of
superconducting cos($2\theta$) quadrupoles, a 6.6-m-long dipole magnet, and a
third superconducting cos($2\theta$) quadrupole.
The detector package included: a pair of vertical
drift chambers to determine the trajectory of a particle; two scintillator
planes to provide the trigger; and a gas Cherenkov detector combined with a
lead-glass electromagnetic calorimeter to separate electrons and pions.
The spectrometer has a solid angle acceptance of 6~msr and a momentum acceptance
of $\pm$~4.5\%.  The spectrometer optics calibration resulted in the following 
resolutions:  6~mm in the vertex position along the beamline, 2$\times 10^{-4}$ in 
relative momentum, 1.5~mrad in the out-of-plane angle and 0.5~mrad in the in-plane 
angle.

During the experiment, two sets of Helmholtz coils were used to align
the $^3$He spin vector either parallel or perpendicular to the beam direction.
This enabled us to measure independently the asymmetries, $A_{\parallel}$ and
$A_{\perp}$, where the subscripts indicate 
the orientation of the $^3$He spin vector with respect to the beam and in the
horizontal lab-frame plane.
The experimental physics asymmetries were calculated by

\begin{equation} 
  A = \frac{1}{P_eP_tf_{\rm{N_{2}}}}\left(\frac{Y^+ - Y^-}
{Y^+ + Y^-}\right),
\end{equation}
where $Y^{\pm} = \frac{N^{\pm}}{Q^{\pm} \cdot LT^{\pm}}$ represents the normalized 
yield for beam helicity $\pm 1$, $N^{\pm}$ is the number of detected
scattered electrons, $Q^{\pm}$ is the accumulated charge, and $LT^{\pm}$ is the
data acquisition live-time. $P_e$ and $P_t$ are the beam and target 
polarizations, respectively, and $f_{\rm{N_{2}}}$ is the dilution factor due 
to the admixture of N$_2$ gas in the target cell.  
The dedicated $\rm{N_{2}}$ reference cell data were used to determine
$f_{\rm{N_{2}}} =$ (95 $\pm$ 2)\%. 
The measured asymmetries were calculated near the quasi-elastic peak for values
of the Bjorken scaling variable $x_\mathrm{B} = Q^2/(2M\omega)$
in the range 0.9~$< x_\mathrm{B} <$~1.1, where
  $M$ is the mass of the nucleon and $\omega = E - E'$,
where $E\hspace{0.1cm}(E')$ is the incident (scattered) electron energy.

Radiative corrections were calculated based on the formalism of Mo and
Tsai~\cite{Mo:1968cg} with the program RADCOR.F~\cite{Slifer_thesis}.
This code was updated to use the peaking approximation of Stein 
{\it et al.}~\cite{Stein:1975yy} and can perform both external and internal
corrections for unpolarized and polarized cross sections.  For the polarized cross sections, 
the relative uncertainty of the radiative corrections was estimated to be 20\% 
and up to 40\%, when extrapolation from the model is involved~\cite{Slifer_thesis}.  The data 
from Ref.~\cite{Slifer:2008re} were used to build a model for the two
helicity states and extrapolated to the kinematics of this paper.  The model cross sections
were then incorporated into the radiative correction procedure.  The size of the corrections 
for the asymmetries varied from 8\% to 14\% across the $x_\mathrm{B}$ acceptance.  Due to 
the assumptions used in building the model and the extrapolation, a conservative relative 
uncertainty of 50\% was chosen for the radiative corrections, which results in a 
relative uncertainty of about 5\% for the corrected asymmetries.  
Since these measurements were done at a moderate $Q^2$ and epsilon near unity,
two-photon effects corrections should be small and thus have been neglected. 

\section{Results}

\begin{figure}[htb]
\begin{center}
\includegraphics[angle=0,width=0.48\textwidth]{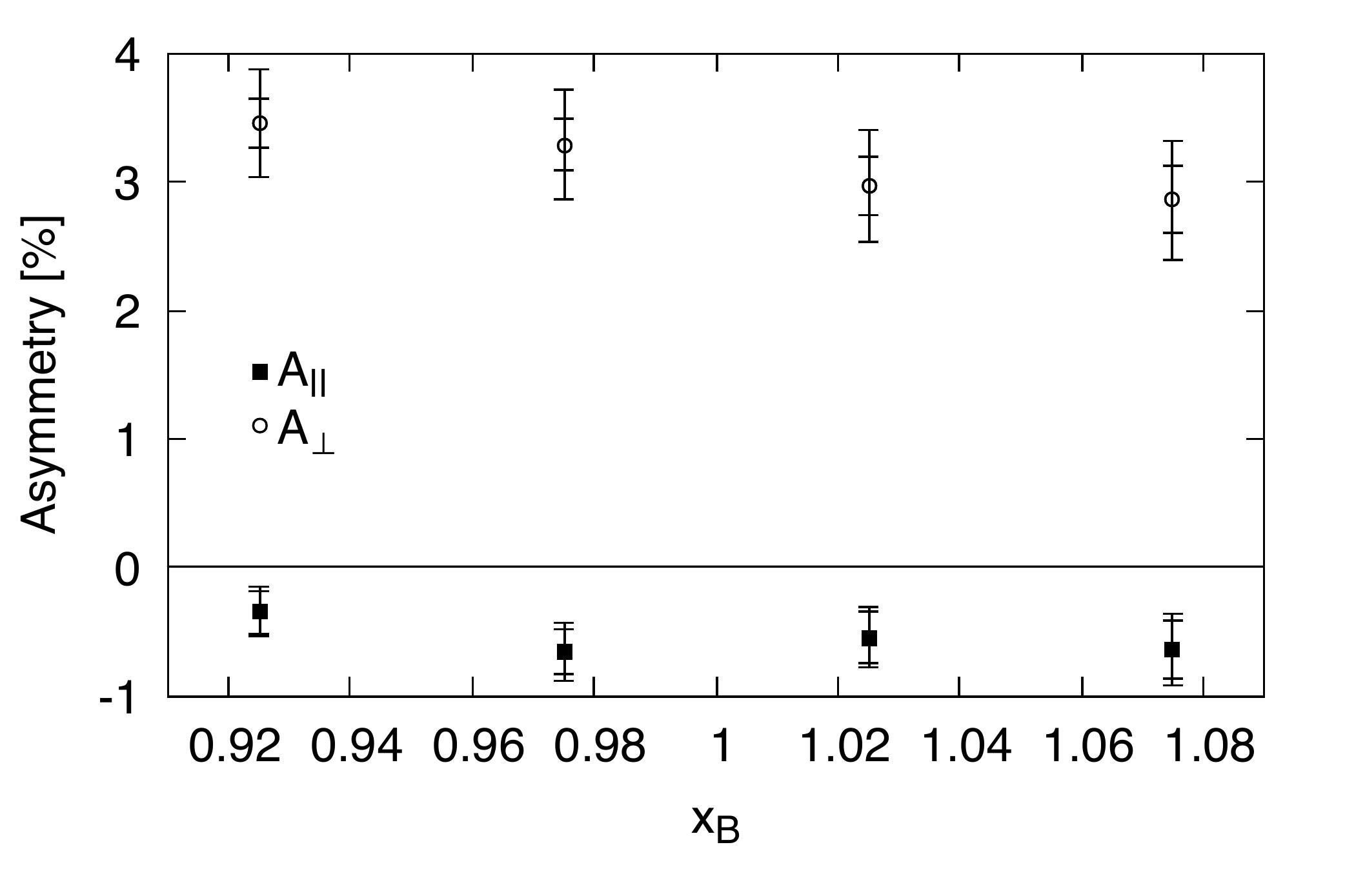}
\caption[Measured asymmetries.]{Inclusive asymmetries from the $^3\overrightarrow{\rm{He}}
(\overrightarrow{e},e')$ reaction with the target spin parallel, $\left(A_{\parallel}\right)$, and
 transverse, $\left(A_{\perp}\right)$, to the electron beam direction asymmetries near 
the quasi-elastic peak versus $x_B$.  The inner (outer) error bars represent the statistical (statistical plus
systematic) uncertainties.
\label{fig:long_rad}}
\end{center}
\end{figure}

The $A_{\parallel}$ and $A_{\perp}$ inclusive $^3$He asymmetries averaged 
over the spectrometer acceptance and after applying radiative corrections are shown 
in Fig.~\ref{fig:long_rad} with their values provided in Table~\ref{tab:radcors}.  
The inner error bars represent the statistical uncertainties; the outer error bars 
show the combined statistical and systematic uncertainties.  For the parallel asymmetry 
$\left(A_\parallel\right)$, the statistical precision overwhelms 
the systematic uncertainty and, hence, the total error bar cannot be easily
distinguished from the statistical error bar. 
The dominant experimental systematic uncertainties for the measured asymmetries
are the uncertainty in the radiative corrections (5\%), the target
polarization (5\%), the beam polarization (3\%) and the dilution factor (2\%),
where all the uncertainties are relative to the asymmetry. The uncertainty due to
inelastic backgrounds was not considered, since the statistical uncertainties
dominate the total uncertainty.
Within the statistical uncertainties, $A_{\parallel}$ is almost constant
across the chosen $x_\mathrm{B}$ range, whereas $A_{\perp}$ exhibits a
slight linear decrease with increasing $x_\mathrm{B}$.

\begin{table}[hb]
\caption{
Parallel $\left(A_{\parallel}\right)$ and transverse $\left(A_{\perp}\right)$ asymmetries near the quasi-elastic peak 
versus $x_\mathrm{B}$.
The format for the asymmetries follows {\it central value} $\pm$ {\it statistical uncertainty}
  $\pm$ {\it systematic uncertainty}.}
\label{tab:radcors}
\begin{center}
\begin{tabular}{lcc}\hline
$x_\mathrm{B}$ &\hspace{0.1cm}$A_{\perp}$ (\%) &\hspace{0.1cm}$A_{\parallel}$ (\%)\\
\hline
0.925       &\hspace{0.1cm} 3.45 $\pm$ 0.19 $\pm$ 0.23      &\hspace{0.1cm} $-$0.35 $\pm$ 0.17 $\pm$ 0.02\\
0.975       &\hspace{0.1cm} 3.29 $\pm$ 0.20 $\pm$ 0.23      &\hspace{0.1cm} $-$0.66 $\pm$ 0.18 $\pm$ 0.05\\
1.025       &\hspace{0.1cm} 2.97 $\pm$ 0.22 $\pm$ 0.21      &\hspace{0.1cm} $-$0.55 $\pm$ 0.20 $\pm$ 0.04\\
1.075       &\hspace{0.1cm} 2.86 $\pm$ 0.26 $\pm$ 0.20      &\hspace{0.1cm} $-$0.64 $\pm$ 0.23 $\pm$ 0.05\\ \hline
\end{tabular}
\end{center}
\end{table}

To relate the measured asymmetries to $G^n_E$ we used the formalism of Donnelly and 
Raskin~\cite{Donnelly:1985ry} for scattering from a free spin-1/2 particle.
The asymmetry for the $^3\overrightarrow{\rm{He}} (\overrightarrow{e},e')$
reaction near the quasi-elastic peak can be written in terms of $^3$He response
functions as the ratio of the spin-averaged ($\Sigma$) and polarization
($\Delta$) cross sections:

\begin{eqnarray}
\lefteqn{A\left(\theta^\ast,\phi^\ast\right) =\frac{\Delta\left(\theta^\ast,\phi^\ast\right)}
{\Sigma\left(\theta^\ast,\phi^\ast\right)}}\nonumber\\
& &=-\frac{\cos\theta^\ast v_{T'}R_{T'}^{\mathrm{^3He}}+
\sin\theta^\ast \cos\phi^\ast v_{TL'}R_{TL'}^\mathrm{^3He}}{v_LR_L^\mathrm{^3He}
+v_TR_T^\mathrm{^3He}},
\label{asy_response_3He}
\end{eqnarray}
where $R_{T'(TL')}^\mathrm{^3He}$ are the $^3$He polarized transverse
(transverse-longitudinal) response functions, $R_{T(L)}^\mathrm{^3He}$ are the
$^3$He unpolarized transverse (longitudinal) response functions, and the $v$'s
are kinematic factors which are independent of beam and target polarizations.
$\theta^\ast$ and $\phi^\ast$ are, respectively, the polar and azimuthal angles 
of the target polarization vector with respect to the three-momentum transfer
$\overrightarrow{q}$.
Thus, asymmetries measured with the target oriented parallel (perpendicular) to
the electron beam correspond to $\theta^* = 56^\circ$ and
$\phi^* = 0^\circ$ $\left(\theta^* = 34^\circ\right.$ and
$\left.\phi^* = 180^\circ\right)$. 

Following the Plane-Wave-Impulse-Approximation (PWIA) calculation by
Kievsky {\it et al.}~\cite{Kievsky:1996gz}, the polarized $^3$He transverse
(transverse-longitudinal) response functions $R_{T'(TL')}^\mathrm{^3He}$ near the
quasi-elastic peak are written as 

\begin{equation}
R_{T'}^\mathrm{^3He}=\frac{Q^2}{2qM}
\left( 2[G_M^p]^2 H_{T'}^p+[G_M^n]^2 H_{T'}^n\right),
\label{RT'}
\end{equation}

\begin{equation}
R_{TL'}^\mathrm{^3He}=-\sqrt{2}
\left( 2G_M^p G_E^p H_{TL'}^p+G_M^nG_E^n H_{TL'}^n\right),
\label{RTL'}
\end{equation}
where the $H_{S}^{p(n)}$ represent the proton (neutron) contribution to
the response functions with $S$ = $T'$ or $TL'$.  
The proton form factors, $G_E^p$ and $G_M^p$, as well as the neutron magnetic form factor $G_M^n$,
were constrained by the world data.
The values of $H_{S}^{n(p)}$ 
were calculated in Ref.~\cite{Kievsky:1996gz} using models for the nucleon polarizations 
and momentum distributions in the
$^3$He nuclei.
These values are almost constant over a wide range of $Q^{2}$.
Thus, by measuring {\it A} for two sets of $\left(\theta^*,\phi^*\right)$ at
electron scattering angles and scattered electron momenta
spanning the acceptance of the RHRS subject to the constraint that
$0.9 < x_\mathrm{B} < 1.1 $, we obtain two linearly independent
equations.
The dependence on the $^3$He unpolarized response functions can be removed by taking the ratio
of $A\left(\theta^\ast,\phi^\ast\right)$ for two sets of  $\left(\theta^*,\phi^*\right)$:
\begin{equation}
\frac{A\left({\theta_2}^\ast,{\phi_2}^\ast\right)}{A\left({\theta_1}^\ast,{\phi_1}^\ast\right)} =
\frac{\cos{\theta_2}^\ast v_{T'}R_{T'}^\mathrm{^3He} +
\sin{\theta_2}^\ast \cos{\phi_2}^\ast v_{TL'}R_{TL'}^\mathrm{^3He}} 
{\cos{\theta_1}^\ast v_{T'}R_{T'}^\mathrm{^3He}+
\sin{\theta_1}^\ast \cos{\phi_1}^\ast v_{TL'}R_{TL'}^\mathrm{^3He}}
\label{Aratio}
\end{equation}
which can be solved for $G_E^n$.

The uncertainties in the ratios of asymmetries were dominated by the statistical uncertainty 
(18.5\%) in the values of $A_{\parallel}$.  In these ratios, the absolute values of corrections 
such as the beam and target polarizations cancel to first order and only their relative changes 
during the measurement contribute to the uncertainty.  We estimate that the uncertainties in the 
asymmetry ratio from the beam and target polarizations are $\simeq$ 3\% and $\simeq$ 1\%, 
respectively.  Similarly, the dilution factors cancel.  The radiative corrections are 
correlated for the two measured asymmetries so that they also mostly cancel in the
ratio.  When the radiative corrections are varied within the uncertainties, we found the ratios of the 
asymmetries change by $\ll$ 1\%; however, due to the assumptions made in their determination, we have 
taken 1\% to be a conservative estimate of the uncertainty from this source.  Finally, the measurement of 
the asymmetries is sensitive to the target polarization angle $\theta^\ast$ that has an uncertainty of 
$\pm$~0.3$^{\circ}$.  This results in a 3\% uncertainty in the ratio.  We estimate the total experimental 
systematic uncertainty to be 4.5\% for the ratios of the asymmetries.  

The discussion up to this point has been based on the PWIA framework.
Corrections to this approximation must be considered.  The effects of 
final state interactions (FSI) were examined and found to decrease significantly 
with increasing $Q^2$~\cite{Pace:1991ct,Benhar:1999ts}.  The PWIA calculation 
mentioned previously~\cite{Kievsky:1996gz} was used in earlier determinations of 
$G_M^n$ in the range $Q^2=0.1$ - $0.6~(\rm{GeV/c})^2$ from measurements of the $A_{T'}$ 
asymmetry made at Jefferson Lab~\cite{Xu:2002xc}.  The effects of FSI were greatly reduced above 
$Q^2$ of 0.5$~(\rm{GeV/c})^2$, and for $Q^{2} \geq 1~(\rm{GeV/c})^2$ FSI corrections are 
expected to fall as $Q^{-4}$.  Corrections for meson-exchange currents (MEC) are 
expected to be negligible at the quasi-elastic peak~\cite{private_sargsian} and to 
decrease exponentially as $Q^2$ increases, based on the observation in Ref.~\cite{Xu:2002xc} 
as well as the calculations of Golak~\cite{Golak:2000nt}.

Within the context of PWIA, inclusion of the off-shell nature of the struck nucleon into the calculation of
the electron-nucleon cross section requires a model of the nucleon current.
In particular, a model for the contribution of the anomalous magnetic moment of the struck nucleon
must be chosen.
The CC1 and CC2 prescriptions of De Forest~\cite{DeForest:1983vc} for off-shell 
cross sections were used to obtain the $^3$He responses.
In these two prescriptions the off-shell effects are incorporated into the electron kinematics
using different approaches as outlined in ref.~\cite{DeForest:1983vc}.
In the CC1 prescription the four-momentum transfer is determined solely by the electron kinematics.
In the CC2 prescription the three-momentum transfer, $\vec{q}$, is determined by the electron kinematics
and the energy transfer from the final energy and initial momentum of the struck nucleon.
It is to be noted that in both cases energy-momentum and current conversation are violated as
in both cases the nucleons are treated as free particles. 
PWIA calculations using these forms provide good agreement with the unpolarized
$^3$He response functions~\cite{Kievsky:1996gz}.  For the polarized responses, both
prescriptions provide essentially the same result for $R_{TL'}$, while the results for 
$R_{T'}$ in general differ less than 2\% over the range of $0.1 \leq Q^2 \leq 2~(\rm{GeV/c})^2$.
Due to these differences, only the results from CC1 were reported in Ref.~\cite{Kievsky:1996gz}.
Other prescriptions are available~\cite{Caballero:1992tt}  but were not considered.

When $x_\mathrm{B}$ is near 1, the struck nucleon is almost at rest
before absorbing the virtual photon.  After absorbing the photon, it has a momentum 
almost equal to that of the virtual photon. In the kinematics of this paper, the 
struck nucleon has a relativistic kinetic energy.  Hence, the inclusion of 
relativistic effects in the theoretical calculations is essential.  The uncertainty 
due to these effects was estimated in Ref.~\cite{private_sargsian}.  A comparison
was made within the Virtual Nucleon and Light Cone approximations, which are 
different treatments of the relativistic motion of bound nucleons as well as
electromagnetic currents.  The difference between the predictions made using these
two approximations was found to be 1.2\% at $Q^2 \simeq 1~(\rm{GeV/c})^2$.

Parameterizations of the three undetermined
electromagnetic form factors were used as inputs to calculate an asymmetry at 
each kinematic point over the measured angular and momentum acceptance of the RHRS.  
For $G_M^n$ the high precision data from Ref.~\cite{Lachniet:2008qf} were used, and
the values for $G_E^p$ and $G_M^p$ were provided by Refs.~\cite{Venkat:2010by,private_arrington}, 
which were extracted after applying the two-photon exchange corrections as done 
in~\cite{Arrington:2007ux}.  The values and uncertainties for the form factors used in the extraction at 
the central value of $Q^2$ are presented in Table~\ref{tab:ffs}.  Taking into account the 
correlations between $G_E^p$ and $G_M^p$, these uncertainties in the 
form factors lead to an uncertainty of 1.4\% in the extracted value of $G_{E}^{n}$.
The extraction of $G_{E}^{n}$ is not limited by the uncertainties on the individual form factors. Examining
Eq.~(\ref{Aratio}) reveals that the proton contributions to the response functions are 
suppressed in $^3$He, and hence, their uncertainties in the extraction of $G_{E}^{n}$ are 
also suppressed. On the other hand, as $Q^2$ increases the uncertainty on 
$G_M^n$ (2.1--2.6\%) becomes important at $Q^2 = 2.6~(\rm{GeV/c})^2$. Finally, the uncertainty on
 $G_{E}^{p}$ grows linearly with $Q^2$ and adds an equal uncertainty in the extraction 
of $G_{E}^{n}$ at this $Q^2$.

\begin{table}[htb]
\caption{ The values and uncertainties for the form factors used in the extraction at $Q^2 = 0.98~(\rm{GeV/c})^2$.  
The column $\delta G_E^n$ provides the contributions to the systematic uncertainty
of $G_E^n$ from the input form factors to Eq.~(\ref{asy_response_3He}).}
\label{tab:ffs}
\begin{center}
\begin{tabular}{lcc}\hline
Form Factor &Value &$\delta G_E^n$\\
\hline
$G_E^p/G_{D}$       &0.9413 $\pm$ 0.0094  &3.0$\times 10^{-4}$ \\
$\mu_pG_M^p/G_{D}$  &1.0456 $\pm$ 0.0104  &2.5$\times 10^{-4}$ \\
$\mu_nG_M^n/G_{D}$  &0.9953 $\pm$ 0.0225  &1.9$\times 10^{-4}$ \\ \hline
\end{tabular}
\end{center}
\end{table}

\begin{figure}[htb]
\begin{center}
\includegraphics[angle=0,width=.5\textwidth]{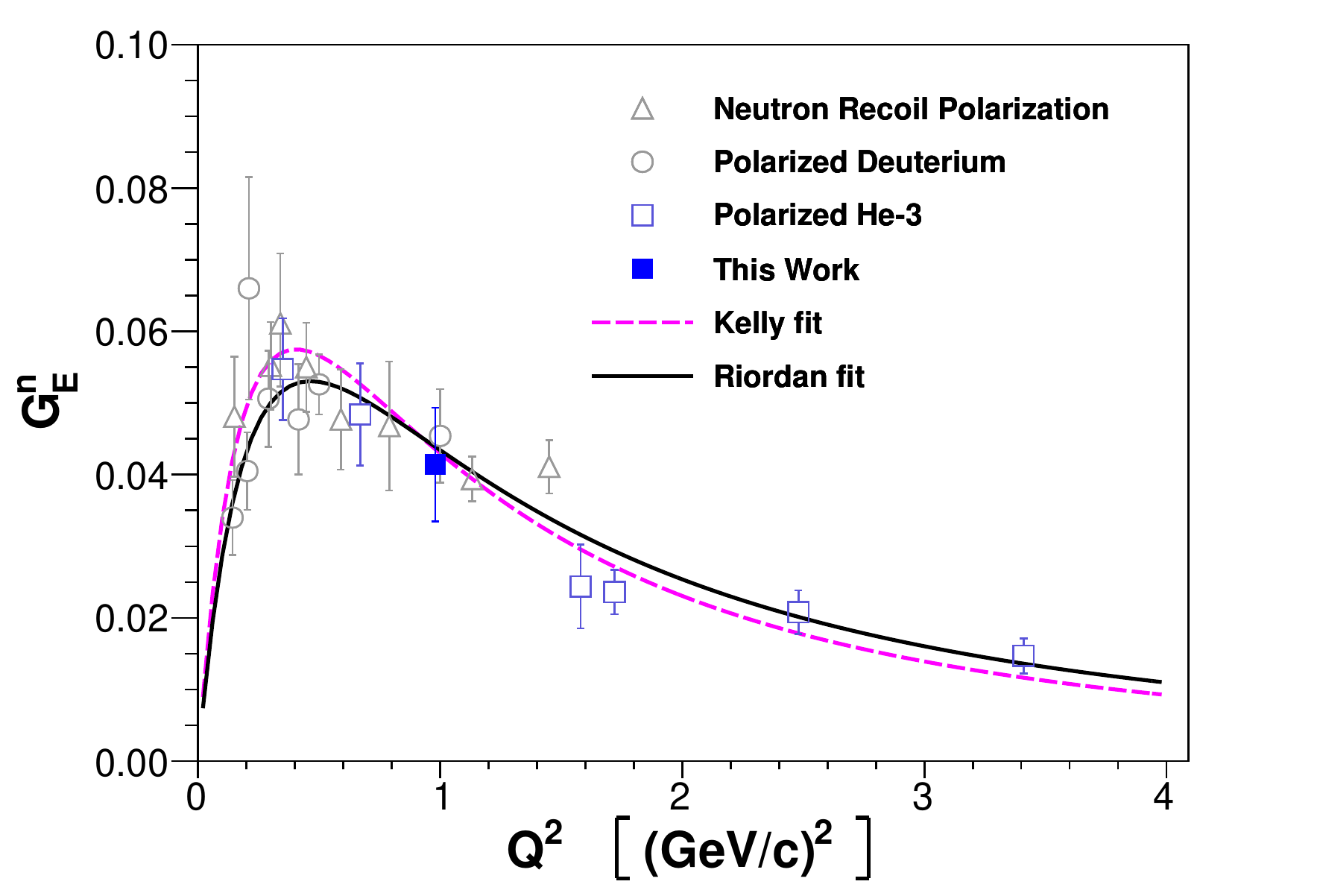}
\caption[$G_E^n$ value from this measurement]{The $G_E^n$ value extracted from this experiment 
(solid square) and selected published data: open triangles~\cite{Herberg:1999ud,Ostrick:1999xa,Glazier:2004ny,Plaster:2005cx}, 
open circles~\cite{Passchier:1999cj,Warren:2003ma,Geis:2008aa}, open 
squares~\cite{Becker:1999tw,Bermuth:2003qh,Riordan:2010id,Schlimme:2013eoz} and parametrizations: 
Riordan {\it et al.}~\cite{Riordan:2010id} and Kelly~\cite{Kelly:2004hm}.  In regards to the 
polarized $^3$He points, the solid square is from the inclusive reaction, whereas the open squares 
represent extracted results from experiments in which the neutron was tagged.  The error bars show 
the statistical and the systematic uncertainties added in quadrature.
\label{gen_plot}}
\end{center}
\end{figure}

Using Eq.~(\ref{Aratio}), the central value of $G_E^n$ was varied 
to fit the calculated ratio of asymmetries to the experimentally measured 
ratios. The value for $G_E^n$ extracted at $Q^2 = 0.98~(\rm{GeV/c})^2$ is

 \begin{equation}
G_E^n=0.0414\pm0.0077\pm0.0022,
\label{final_result2}
\end{equation}
where the first (second) uncertainty is statistical (systematic).
In Fig.~\ref{gen_plot}, the present result for $G_E^n$ is shown as the solid 
square along with selected world data and parametrizations. The extracted 
result is consistent with the world data, showing the feasibility of this 
method for values of $Q^{2}$ larger than $0.8~(\rm{GeV/c})^2$.
It should be noted that the present data were acquired in only 2.5 days 
of running. As the extraction of $G_E^n$ was not the principal focus of 
the measurements, the running time was divided evenly between the two target 
polarization orientations: parallel to the electron beam and perpendicular 
to the beam. Had the division of running time been optimized, with 90\% (10\%) 
of the time allocated to the parallel (perpendicular) orientation, the 
statistical uncertainty on $G_E^n$ would be reduced from 0.0077 to 0.0026. 

\section{Summary}

In summary, an extraction of $G_E^n$ from inclusive polarized
$^3\overrightarrow{\rm{He}}(\overrightarrow{e},e')$ quasi-elastic asymmetry measurements was
presented.
This  method of forming the ratio of inclusive asymmetries provides an important
independent check of other measurements and has several advantages.
Firstly, the systematic uncertainties associated with neutron
detection~\cite{Riordan:2010id} are avoided. 
Secondly, the sensitivity to certain unavoidable systematic errors (beam and
target polarizations, dilution factors and radiative corrections) are greatly
reduced due to first-order cancellations in the ratio of asymmetries.
The final result at $Q^2 = 0.98~(\rm{GeV}/c)^2$ of $G_E^n=0.0414\pm0.0077\pm0.0019$
was found to be consistent with other extraction techniques.   
This is in contrast to the proton, where at this same $Q^2$, systematic 
differences between form factor extraction techniques were revealed~\cite{Jones:1999rz}.

We thank the Jefferson Lab Physics and Accelerator Divisions. This work was
supported in part by the U.S. National Science Foundation and by the U.S. 
Department of Energy.  It is supported by DOE contract No. DE-AC05-06OR23177, 
under which Jefferson Science Associates (JSA), LLC, operates the Thomas 
Jefferson National Accelerator Facility.

\bibliography{inspire}

\begin{thebibliography}{41}%
\makeatletter
\providecommand \@ifxundefined [1]{%
 \@ifx{#1\undefined}
}%
\providecommand \@ifnum [1]{%
 \ifnum #1\expandafter \@firstoftwo
 \else \expandafter \@secondoftwo
 \fi
}%
\providecommand \@ifx [1]{%
 \ifx #1\expandafter \@firstoftwo
 \else \expandafter \@secondoftwo
 \fi
}%
\providecommand \natexlab [1]{#1}%
\providecommand \enquote  [1]{``#1''}%
\providecommand \bibnamefont  [1]{#1}%
\providecommand \bibfnamefont [1]{#1}%
\providecommand \citenamefont [1]{#1}%
\providecommand \href@noop [0]{\@secondoftwo}%
\providecommand \href [0]{\begingroup \@sanitize@url \@href}%
\providecommand \@href[1]{\@@startlink{#1}\@@href}%
\providecommand \@@href[1]{\endgroup#1\@@endlink}%
\providecommand \@sanitize@url [0]{\catcode `\\12\catcode `\$12\catcode
  `\&12\catcode `\#12\catcode `\^12\catcode `\_12\catcode `\%12\relax}%
\providecommand \@@startlink[1]{}%
\providecommand \@@endlink[0]{}%
\providecommand \url  [0]{\begingroup\@sanitize@url \@url }%
\providecommand \@url [1]{\endgroup\@href {#1}{\urlprefix }}%
\providecommand \urlprefix  [0]{URL }%
\providecommand \Eprint [0]{\href }%
\providecommand \doibase [0]{http://dx.doi.org/}%
\providecommand \selectlanguage [0]{\@gobble}%
\providecommand \bibinfo  [0]{\@secondoftwo}%
\providecommand \bibfield  [0]{\@secondoftwo}%
\providecommand \translation [1]{[#1]}%
\providecommand \BibitemOpen [0]{}%
\providecommand \bibitemStop [0]{}%
\providecommand \bibitemNoStop [0]{.\EOS\space}%
\providecommand \EOS [0]{\spacefactor3000\relax}%
\providecommand \BibitemShut  [1]{\csname bibitem#1\endcsname}%
\let\auto@bib@innerbib\@empty
\bibitem [{\citenamefont {Galster}\ \emph {et~al.}(1971)\citenamefont
  {Galster}, \citenamefont {Klein}, \citenamefont {Moritz}, \citenamefont
  {Schmidt}, \citenamefont {Wegener},\ and\ \citenamefont
  {Bleckwenn}}]{Galster:1971kv}%
  \BibitemOpen
  \bibfield  {author} {\bibinfo {author} {\bibfnamefont {S.}~\bibnamefont
  {Galster}}, \bibinfo {author} {\bibfnamefont {H.}~\bibnamefont {Klein}},
  \bibinfo {author} {\bibfnamefont {J.}~\bibnamefont {Moritz}}, \bibinfo
  {author} {\bibfnamefont {K.~H.}\ \bibnamefont {Schmidt}}, \bibinfo {author}
  {\bibfnamefont {D.}~\bibnamefont {Wegener}}, \ and\ \bibinfo {author}
  {\bibfnamefont {J.}~\bibnamefont {Bleckwenn}},\ }\href {\doibase
  10.1016/0550-3213(71)90068-X} {\bibfield  {journal} {\bibinfo  {journal}
  {Nucl. Phys.}\ }\textbf {\bibinfo {volume} {B32}},\ \bibinfo {pages} {221}
  (\bibinfo {year} {1971})}\BibitemShut {NoStop}%
\bibitem [{\citenamefont {Platchkov}\ \emph {et~al.}(1990)\citenamefont
  {Platchkov} \emph {et~al.}}]{Platchkov:1989ch}%
  \BibitemOpen
  \bibfield  {author} {\bibinfo {author} {\bibfnamefont {S.}~\bibnamefont
  {Platchkov}} \emph {et~al.},\ }\href {\doibase 10.1016/0375-9474(90)90358-S}
  {\bibfield  {journal} {\bibinfo  {journal} {Nucl. Phys.}\ }\textbf {\bibinfo
  {volume} {A510}},\ \bibinfo {pages} {740} (\bibinfo {year}
  {1990})}\BibitemShut {NoStop}%
\bibitem [{\citenamefont {Passchier}\ \emph {et~al.}(1999)\citenamefont
  {Passchier} \emph {et~al.}}]{Passchier:1999cj}%
  \BibitemOpen
  \bibfield  {author} {\bibinfo {author} {\bibfnamefont {I.}~\bibnamefont
  {Passchier}} \emph {et~al.},\ }\href {\doibase 10.1103/PhysRevLett.82.4988}
  {\bibfield  {journal} {\bibinfo  {journal} {Phys. Rev. Lett.}\ }\textbf
  {\bibinfo {volume} {82}},\ \bibinfo {pages} {4988} (\bibinfo {year}
  {1999})},\ \Eprint {http://arxiv.org/abs/nucl-ex/9907012}
  {arXiv:nucl-ex/9907012 [nucl-ex]} \BibitemShut {NoStop}%
\bibitem [{\citenamefont {Warren}\ \emph {et~al.}(2004)\citenamefont {Warren}
  \emph {et~al.}}]{Warren:2003ma}%
  \BibitemOpen
  \bibfield  {author} {\bibinfo {author} {\bibfnamefont {G.}~\bibnamefont
  {Warren}} \emph {et~al.} (\bibinfo {collaboration} {Jefferson Lab E93-026}),\
  }\href {\doibase 10.1103/PhysRevLett.92.042301} {\bibfield  {journal}
  {\bibinfo  {journal} {Phys. Rev. Lett.}\ }\textbf {\bibinfo {volume} {92}},\
  \bibinfo {pages} {042301} (\bibinfo {year} {2004})},\ \Eprint
  {http://arxiv.org/abs/nucl-ex/0308021} {arXiv:nucl-ex/0308021 [nucl-ex]}
  \BibitemShut {NoStop}%
\bibitem [{\citenamefont {Geis}\ \emph {et~al.}(2008)\citenamefont {Geis} \emph
  {et~al.}}]{Geis:2008aa}%
  \BibitemOpen
  \bibfield  {author} {\bibinfo {author} {\bibfnamefont {E.}~\bibnamefont
  {Geis}} \emph {et~al.} (\bibinfo {collaboration} {BLAST}),\ }\href {\doibase
  10.1103/PhysRevLett.101.042501} {\bibfield  {journal} {\bibinfo  {journal}
  {Phys. Rev. Lett.}\ }\textbf {\bibinfo {volume} {101}},\ \bibinfo {pages}
  {042501} (\bibinfo {year} {2008})},\ \Eprint {http://arxiv.org/abs/0803.3827}
  {arXiv:0803.3827 [nucl-ex]} \BibitemShut {NoStop}%
\bibitem [{\citenamefont {Becker}\ \emph {et~al.}(1999)\citenamefont {Becker}
  \emph {et~al.}}]{Becker:1999tw}%
  \BibitemOpen
  \bibfield  {author} {\bibinfo {author} {\bibfnamefont {J.}~\bibnamefont
  {Becker}} \emph {et~al.},\ }\href {\doibase 10.1007/s100500050351} {\bibfield
   {journal} {\bibinfo  {journal} {Eur. Phys. J.}\ }\textbf {\bibinfo {volume}
  {A6}},\ \bibinfo {pages} {329} (\bibinfo {year} {1999})}\BibitemShut
  {NoStop}%
\bibitem [{\citenamefont {Bermuth}\ \emph {et~al.}(2003)\citenamefont {Bermuth}
  \emph {et~al.}}]{Bermuth:2003qh}%
  \BibitemOpen
  \bibfield  {author} {\bibinfo {author} {\bibfnamefont {J.}~\bibnamefont
  {Bermuth}} \emph {et~al.},\ }\href {\doibase 10.1016/S0370-2693(03)00725-1}
  {\bibfield  {journal} {\bibinfo  {journal} {Phys. Lett.}\ }\textbf {\bibinfo
  {volume} {B564}},\ \bibinfo {pages} {199} (\bibinfo {year} {2003})},\ \Eprint
  {http://arxiv.org/abs/nucl-ex/0303015} {arXiv:nucl-ex/0303015 [nucl-ex]}
  \BibitemShut {NoStop}%
\bibitem [{\citenamefont {Riordan}\ \emph {et~al.}(2010)\citenamefont {Riordan}
  \emph {et~al.}}]{Riordan:2010id}%
  \BibitemOpen
  \bibfield  {author} {\bibinfo {author} {\bibfnamefont {S.}~\bibnamefont
  {Riordan}} \emph {et~al.},\ }\href {\doibase 10.1103/PhysRevLett.105.262302}
  {\bibfield  {journal} {\bibinfo  {journal} {Phys. Rev. Lett.}\ }\textbf
  {\bibinfo {volume} {105}},\ \bibinfo {pages} {262302} (\bibinfo {year}
  {2010})},\ \Eprint {http://arxiv.org/abs/1008.1738} {arXiv:1008.1738
  [nucl-ex]} \BibitemShut {NoStop}%
\bibitem [{\citenamefont {Schlimme}\ \emph {et~al.}(2013)\citenamefont
  {Schlimme} \emph {et~al.}}]{Schlimme:2013eoz}%
  \BibitemOpen
  \bibfield  {author} {\bibinfo {author} {\bibfnamefont {B.~S.}\ \bibnamefont
  {Schlimme}} \emph {et~al.},\ }\href {\doibase 10.1103/PhysRevLett.111.132504}
  {\bibfield  {journal} {\bibinfo  {journal} {Phys. Rev. Lett.}\ }\textbf
  {\bibinfo {volume} {111}},\ \bibinfo {pages} {132504} (\bibinfo {year}
  {2013})},\ \Eprint {http://arxiv.org/abs/1307.7361} {arXiv:1307.7361
  [nucl-ex]} \BibitemShut {NoStop}%
\bibitem [{\citenamefont {Herberg}\ \emph {et~al.}(1999)\citenamefont {Herberg}
  \emph {et~al.}}]{Herberg:1999ud}%
  \BibitemOpen
  \bibfield  {author} {\bibinfo {author} {\bibfnamefont {C.}~\bibnamefont
  {Herberg}} \emph {et~al.},\ }\href {\doibase 10.1007/s100500050268}
  {\bibfield  {journal} {\bibinfo  {journal} {Eur. Phys. J.}\ }\textbf
  {\bibinfo {volume} {A5}},\ \bibinfo {pages} {131} (\bibinfo {year}
  {1999})}\BibitemShut {NoStop}%
\bibitem [{\citenamefont {Ostrick}\ \emph {et~al.}(1999)\citenamefont {Ostrick}
  \emph {et~al.}}]{Ostrick:1999xa}%
  \BibitemOpen
  \bibfield  {author} {\bibinfo {author} {\bibfnamefont {M.}~\bibnamefont
  {Ostrick}} \emph {et~al.},\ }\href {\doibase 10.1103/PhysRevLett.83.276}
  {\bibfield  {journal} {\bibinfo  {journal} {Phys. Rev. Lett.}\ }\textbf
  {\bibinfo {volume} {83}},\ \bibinfo {pages} {276} (\bibinfo {year}
  {1999})}\BibitemShut {NoStop}%
\bibitem [{\citenamefont {Glazier}\ \emph {et~al.}(2005)\citenamefont {Glazier}
  \emph {et~al.}}]{Glazier:2004ny}%
  \BibitemOpen
  \bibfield  {author} {\bibinfo {author} {\bibfnamefont {D.~I.}\ \bibnamefont
  {Glazier}} \emph {et~al.},\ }\href {\doibase 10.1140/epja/i2004-10115-8}
  {\bibfield  {journal} {\bibinfo  {journal} {Eur. Phys. J.}\ }\textbf
  {\bibinfo {volume} {A24}},\ \bibinfo {pages} {101} (\bibinfo {year}
  {2005})},\ \Eprint {http://arxiv.org/abs/nucl-ex/0410026}
  {arXiv:nucl-ex/0410026 [nucl-ex]} \BibitemShut {NoStop}%
\bibitem [{\citenamefont {Plaster}\ \emph {et~al.}(2006)\citenamefont {Plaster}
  \emph {et~al.}}]{Plaster:2005cx}%
  \BibitemOpen
  \bibfield  {author} {\bibinfo {author} {\bibfnamefont {B.}~\bibnamefont
  {Plaster}} \emph {et~al.} (\bibinfo {collaboration} {Jefferson Laboratory
  E93-038}),\ }\href {\doibase 10.1103/PhysRevC.73.025205} {\bibfield
  {journal} {\bibinfo  {journal} {Phys. Rev.}\ }\textbf {\bibinfo {volume}
  {C73}},\ \bibinfo {pages} {025205} (\bibinfo {year} {2006})},\ \Eprint
  {http://arxiv.org/abs/nucl-ex/0511025} {arXiv:nucl-ex/0511025 [nucl-ex]}
  \BibitemShut {NoStop}%
\bibitem [{\citenamefont {Jones-Woodward}\ \emph {et~al.}(1991)\citenamefont
  {Jones-Woodward} \emph {et~al.}}]{JonesWoodward:1991ih}%
  \BibitemOpen
  \bibfield  {author} {\bibinfo {author} {\bibfnamefont {C.~E.}\ \bibnamefont
  {Jones-Woodward}} \emph {et~al.},\ }\href {\doibase 10.1103/PhysRevC.44.R571}
  {\bibfield  {journal} {\bibinfo  {journal} {Phys. Rev.}\ }\textbf {\bibinfo
  {volume} {C44}},\ \bibinfo {pages} {R571} (\bibinfo {year}
  {1991})}\BibitemShut {NoStop}%
\bibitem [{\citenamefont {Thompson}\ \emph {et~al.}(1992)\citenamefont
  {Thompson} \emph {et~al.}}]{Thompson:1992ci}%
  \BibitemOpen
  \bibfield  {author} {\bibinfo {author} {\bibfnamefont {A.~K.}\ \bibnamefont
  {Thompson}} \emph {et~al.},\ }\href {\doibase 10.1103/PhysRevLett.68.2901}
  {\bibfield  {journal} {\bibinfo  {journal} {Phys. Rev. Lett.}\ }\textbf
  {\bibinfo {volume} {68}},\ \bibinfo {pages} {2901} (\bibinfo {year}
  {1992})}\BibitemShut {NoStop}%
\bibitem [{\citenamefont {Hansen}\ \emph {et~al.}(1995)\citenamefont {Hansen}
  \emph {et~al.}}]{Hansen:1994ba}%
  \BibitemOpen
  \bibfield  {author} {\bibinfo {author} {\bibfnamefont {J.~O.}\ \bibnamefont
  {Hansen}} \emph {et~al.},\ }\href {\doibase 10.1103/PhysRevLett.74.654}
  {\bibfield  {journal} {\bibinfo  {journal} {Phys. Rev. Lett.}\ }\textbf
  {\bibinfo {volume} {74}},\ \bibinfo {pages} {654} (\bibinfo {year}
  {1995})}\BibitemShut {NoStop}%
\bibitem [{\citenamefont {Androic}\ \emph {et~al.}(2011)\citenamefont {Androic}
  \emph {et~al.}}]{Androic:2011rha}%
  \BibitemOpen
  \bibfield  {author} {\bibinfo {author} {\bibfnamefont {D.}~\bibnamefont
  {Androic}} \emph {et~al.} (\bibinfo {collaboration} {G0}),\ }\href {\doibase
  10.1016/j.nima.2011.04.031} {\bibfield  {journal} {\bibinfo  {journal} {Nucl.
  Instrum. Meth.}\ }\textbf {\bibinfo {volume} {A646}},\ \bibinfo {pages} {59}
  (\bibinfo {year} {2011})},\ \Eprint {http://arxiv.org/abs/1103.0761}
  {arXiv:1103.0761 [nucl-ex]} \BibitemShut {NoStop}%
\bibitem [{\citenamefont {Jin}(2011)}]{JinGe_thesis}%
  \BibitemOpen
  \bibfield  {author} {\bibinfo {author} {\bibfnamefont {G.}~\bibnamefont
  {Jin}},\ }\href
  {https://misportal.jlab.org/ul/publications/view_pub.cfm?pub_id=14856}
  {\bibfield  {journal} {\bibinfo  {journal} {Ph.D. thesis, University of
  Virginia}\ } (\bibinfo {year} {2011})}\BibitemShut {NoStop}%
\bibitem [{\citenamefont {Babcock}\ \emph {et~al.}(2003)\citenamefont
  {Babcock}, \citenamefont {Nelson}, \citenamefont {Kadlecek}, \citenamefont
  {Driehuys}, \citenamefont {Anderson}, \citenamefont {Hersman},\ and\
  \citenamefont {Walker}}]{Babcock:2003zz}%
  \BibitemOpen
  \bibfield  {author} {\bibinfo {author} {\bibfnamefont {E.}~\bibnamefont
  {Babcock}}, \bibinfo {author} {\bibfnamefont {I.}~\bibnamefont {Nelson}},
  \bibinfo {author} {\bibfnamefont {S.}~\bibnamefont {Kadlecek}}, \bibinfo
  {author} {\bibfnamefont {B.}~\bibnamefont {Driehuys}}, \bibinfo {author}
  {\bibfnamefont {L.~W.}\ \bibnamefont {Anderson}}, \bibinfo {author}
  {\bibfnamefont {F.~W.}\ \bibnamefont {Hersman}}, \ and\ \bibinfo {author}
  {\bibfnamefont {T.~G.}\ \bibnamefont {Walker}},\ }\href {\doibase
  10.1103/PhysRevLett.91.123003} {\bibfield  {journal} {\bibinfo  {journal}
  {Phys. Rev. Lett.}\ }\textbf {\bibinfo {volume} {91}},\ \bibinfo {pages}
  {123003} (\bibinfo {year} {2003})}\BibitemShut {NoStop}%
\bibitem [{\citenamefont {Singh}\ \emph {et~al.}(2015)\citenamefont {Singh},
  \citenamefont {Dolph}, \citenamefont {Tobias}, \citenamefont {Averett},
  \citenamefont {Kelleher}, \citenamefont {Mooney}, \citenamefont {Nelyubin},
  \citenamefont {Wang}, \citenamefont {Zheng},\ and\ \citenamefont
  {Cates}}]{Singh:2013nja}%
  \BibitemOpen
  \bibfield  {author} {\bibinfo {author} {\bibfnamefont {J.}~\bibnamefont
  {Singh}}, \bibinfo {author} {\bibfnamefont {P.~A.~M.}\ \bibnamefont {Dolph}},
  \bibinfo {author} {\bibfnamefont {W.~A.}\ \bibnamefont {Tobias}}, \bibinfo
  {author} {\bibfnamefont {T.~D.}\ \bibnamefont {Averett}}, \bibinfo {author}
  {\bibfnamefont {A.}~\bibnamefont {Kelleher}}, \bibinfo {author}
  {\bibfnamefont {K.~E.}\ \bibnamefont {Mooney}}, \bibinfo {author}
  {\bibfnamefont {V.~V.}\ \bibnamefont {Nelyubin}}, \bibinfo {author}
  {\bibfnamefont {Y.}~\bibnamefont {Wang}}, \bibinfo {author} {\bibfnamefont
  {Y.}~\bibnamefont {Zheng}}, \ and\ \bibinfo {author} {\bibfnamefont {G.~D.}\
  \bibnamefont {Cates}},\ }\href {\doibase 10.1103/PhysRevC.91.055205}
  {\bibfield  {journal} {\bibinfo  {journal} {Phys. Rev.}\ }\textbf {\bibinfo
  {volume} {C91}},\ \bibinfo {pages} {055205} (\bibinfo {year} {2015})},\
  \Eprint {http://arxiv.org/abs/1309.4004} {arXiv:1309.4004 [physics.atom-ph]}
  \BibitemShut {NoStop}%
\bibitem [{\citenamefont {Romalis}\ \emph {et~al.}(1998)\citenamefont {Romalis}
  \emph {et~al.}}]{Romalis:1998ik}%
  \BibitemOpen
  \bibfield  {author} {\bibinfo {author} {\bibfnamefont {M.~V.}\ \bibnamefont
  {Romalis}} \emph {et~al.},\ }\bibfield  {booktitle} {\emph {\bibinfo
  {booktitle} {{Polarized He-3 beams and gas targets and their application.
  Proceedings, 7th RCNP International Workshop, HELION'97, Kobe, Japan, January
  20-24, 1997}}},\ }\href {\doibase 10.1016/S0168-9002(97)00847-4} {\bibfield
  {journal} {\bibinfo  {journal} {Nucl. Instrum. Meth.}\ }\textbf {\bibinfo
  {volume} {A402}},\ \bibinfo {pages} {260} (\bibinfo {year}
  {1998})}\BibitemShut {NoStop}%
\bibitem [{\citenamefont {Alcorn}\ \emph {et~al.}(2004)\citenamefont {Alcorn}
  \emph {et~al.}}]{Alcorn:2004sb}%
  \BibitemOpen
  \bibfield  {author} {\bibinfo {author} {\bibfnamefont {J.}~\bibnamefont
  {Alcorn}} \emph {et~al.},\ }\href {\doibase 10.1016/j.nima.2003.11.415}
  {\bibfield  {journal} {\bibinfo  {journal} {Nucl. Instrum. Meth.}\ }\textbf
  {\bibinfo {volume} {A522}},\ \bibinfo {pages} {294} (\bibinfo {year}
  {2004})}\BibitemShut {NoStop}%
\bibitem [{\citenamefont {Mo}\ and\ \citenamefont {Tsai}(1969)}]{Mo:1968cg}%
  \BibitemOpen
  \bibfield  {author} {\bibinfo {author} {\bibfnamefont {L.~W.}\ \bibnamefont
  {Mo}}\ and\ \bibinfo {author} {\bibfnamefont {Y.-S.}\ \bibnamefont {Tsai}},\
  }\href {\doibase 10.1103/RevModPhys.41.205} {\bibfield  {journal} {\bibinfo
  {journal} {Rev. Mod. Phys.}\ }\textbf {\bibinfo {volume} {41}},\ \bibinfo
  {pages} {205} (\bibinfo {year} {1969})}\BibitemShut {NoStop}%
\bibitem [{\citenamefont {Slifer}(2004)}]{Slifer_thesis}%
  \BibitemOpen
  \bibfield  {author} {\bibinfo {author} {\bibfnamefont {K.}~\bibnamefont
  {Slifer}},\ }\href@noop {} {\bibfield  {journal} {\bibinfo  {journal} {Ph.D.
  thesis, Temple University}\ } (\bibinfo {year} {2004})}\BibitemShut {NoStop}%
\bibitem [{\citenamefont {Stein}\ \emph {et~al.}(1975)\citenamefont {Stein},
  \citenamefont {Atwood}, \citenamefont {Bloom}, \citenamefont {Cottrell},
  \citenamefont {DeStaebler}, \citenamefont {Jordan}, \citenamefont {Piel},
  \citenamefont {Prescott}, \citenamefont {Siemann},\ and\ \citenamefont
  {Taylor}}]{Stein:1975yy}%
  \BibitemOpen
  \bibfield  {author} {\bibinfo {author} {\bibfnamefont {S.}~\bibnamefont
  {Stein}}, \bibinfo {author} {\bibfnamefont {W.~B.}\ \bibnamefont {Atwood}},
  \bibinfo {author} {\bibfnamefont {E.~D.}\ \bibnamefont {Bloom}}, \bibinfo
  {author} {\bibfnamefont {R.~L.}\ \bibnamefont {Cottrell}}, \bibinfo {author}
  {\bibfnamefont {H.~C.}\ \bibnamefont {DeStaebler}}, \bibinfo {author}
  {\bibfnamefont {C.~L.}\ \bibnamefont {Jordan}}, \bibinfo {author}
  {\bibfnamefont {H.}~\bibnamefont {Piel}}, \bibinfo {author} {\bibfnamefont
  {C.~Y.}\ \bibnamefont {Prescott}}, \bibinfo {author} {\bibfnamefont
  {R.}~\bibnamefont {Siemann}}, \ and\ \bibinfo {author} {\bibfnamefont
  {R.~E.}\ \bibnamefont {Taylor}},\ }\href {\doibase 10.1103/PhysRevD.12.1884}
  {\bibfield  {journal} {\bibinfo  {journal} {Phys. Rev.}\ }\textbf {\bibinfo
  {volume} {D12}},\ \bibinfo {pages} {1884} (\bibinfo {year}
  {1975})}\BibitemShut {NoStop}%
\bibitem [{\citenamefont {Slifer}\ \emph {et~al.}(2008)\citenamefont {Slifer}
  \emph {et~al.}}]{Slifer:2008re}%
  \BibitemOpen
  \bibfield  {author} {\bibinfo {author} {\bibfnamefont {K.}~\bibnamefont
  {Slifer}} \emph {et~al.} (\bibinfo {collaboration} {E94010}),\ }\href
  {\doibase 10.1103/PhysRevLett.101.022303} {\bibfield  {journal} {\bibinfo
  {journal} {Phys. Rev. Lett.}\ }\textbf {\bibinfo {volume} {101}},\ \bibinfo
  {pages} {022303} (\bibinfo {year} {2008})},\ \Eprint
  {http://arxiv.org/abs/0803.2267} {arXiv:0803.2267 [nucl-ex]} \BibitemShut
  {NoStop}%
\bibitem [{\citenamefont {Donnelly}\ and\ \citenamefont
  {Raskin}(1986)}]{Donnelly:1985ry}%
  \BibitemOpen
  \bibfield  {author} {\bibinfo {author} {\bibfnamefont {T.~W.}\ \bibnamefont
  {Donnelly}}\ and\ \bibinfo {author} {\bibfnamefont {A.~S.}\ \bibnamefont
  {Raskin}},\ }\href {\doibase 10.1016/0003-4916(86)90173-9} {\bibfield
  {journal} {\bibinfo  {journal} {Annals Phys.}\ }\textbf {\bibinfo {volume}
  {169}},\ \bibinfo {pages} {247} (\bibinfo {year} {1986})}\BibitemShut
  {NoStop}%
\bibitem [{\citenamefont {Kievsky}\ \emph {et~al.}(1997)\citenamefont
  {Kievsky}, \citenamefont {Pace}, \citenamefont {Salme},\ and\ \citenamefont
  {Viviani}}]{Kievsky:1996gz}%
  \BibitemOpen
  \bibfield  {author} {\bibinfo {author} {\bibfnamefont {A.}~\bibnamefont
  {Kievsky}}, \bibinfo {author} {\bibfnamefont {E.}~\bibnamefont {Pace}},
  \bibinfo {author} {\bibfnamefont {G.}~\bibnamefont {Salme}}, \ and\ \bibinfo
  {author} {\bibfnamefont {M.}~\bibnamefont {Viviani}},\ }\href {\doibase
  10.1103/PhysRevC.56.64} {\bibfield  {journal} {\bibinfo  {journal} {Phys.
  Rev.}\ }\textbf {\bibinfo {volume} {C56}},\ \bibinfo {pages} {64} (\bibinfo
  {year} {1997})},\ \Eprint {http://arxiv.org/abs/nucl-th/9704050}
  {arXiv:nucl-th/9704050 [nucl-th]} \BibitemShut {NoStop}%
\bibitem [{\citenamefont {Pace}\ \emph {et~al.}(1991)\citenamefont {Pace},
  \citenamefont {Salme},\ and\ \citenamefont {West}}]{Pace:1991ct}%
  \BibitemOpen
  \bibfield  {author} {\bibinfo {author} {\bibfnamefont {E.}~\bibnamefont
  {Pace}}, \bibinfo {author} {\bibfnamefont {G.}~\bibnamefont {Salme}}, \ and\
  \bibinfo {author} {\bibfnamefont {G.~B.}\ \bibnamefont {West}},\ }\href
  {\doibase 10.1016/0370-2693(91)91672-I} {\bibfield  {journal} {\bibinfo
  {journal} {Phys. Lett.}\ }\textbf {\bibinfo {volume} {B273}},\ \bibinfo
  {pages} {205} (\bibinfo {year} {1991})}\BibitemShut {NoStop}%
\bibitem [{\citenamefont {Benhar}(1999)}]{Benhar:1999ts}%
  \BibitemOpen
  \bibfield  {author} {\bibinfo {author} {\bibfnamefont {O.}~\bibnamefont
  {Benhar}},\ }\href {\doibase 10.1103/PhysRevLett.83.3130} {\bibfield
  {journal} {\bibinfo  {journal} {Phys. Rev. Lett.}\ }\textbf {\bibinfo
  {volume} {83}},\ \bibinfo {pages} {3130} (\bibinfo {year} {1999})},\ \Eprint
  {http://arxiv.org/abs/nucl-th/9908086} {arXiv:nucl-th/9908086 [nucl-th]}
  \BibitemShut {NoStop}%
\bibitem [{\citenamefont {Xu}\ \emph {et~al.}(2003)\citenamefont {Xu} \emph
  {et~al.}}]{Xu:2002xc}%
  \BibitemOpen
  \bibfield  {author} {\bibinfo {author} {\bibfnamefont {W.}~\bibnamefont {Xu}}
  \emph {et~al.} (\bibinfo {collaboration} {Jefferson Lab E95-001}),\ }\href
  {\doibase 10.1103/PhysRevC.67.012201} {\bibfield  {journal} {\bibinfo
  {journal} {Phys. Rev.}\ }\textbf {\bibinfo {volume} {C67}},\ \bibinfo {pages}
  {012201} (\bibinfo {year} {2003})},\ \Eprint
  {http://arxiv.org/abs/nucl-ex/0208007} {arXiv:nucl-ex/0208007 [nucl-ex]}
  \BibitemShut {NoStop}%
\bibitem [{\citenamefont {Sargsian}(2012)}]{private_sargsian}%
  \BibitemOpen
  \bibfield  {author} {\bibinfo {author} {\bibfnamefont {M.}~\bibnamefont
  {Sargsian}},\ }\href@noop {} {\bibfield  {journal} {\bibinfo  {journal}
  {private communication}\ } (\bibinfo {year} {2012})}\BibitemShut {NoStop}%
\bibitem [{\citenamefont {Golak}\ \emph {et~al.}(2001)\citenamefont {Golak},
  \citenamefont {Ziemer}, \citenamefont {Kamada}, \citenamefont {Witala},\ and\
  \citenamefont {Gloeckle}}]{Golak:2000nt}%
  \BibitemOpen
  \bibfield  {author} {\bibinfo {author} {\bibfnamefont {J.}~\bibnamefont
  {Golak}}, \bibinfo {author} {\bibfnamefont {G.}~\bibnamefont {Ziemer}},
  \bibinfo {author} {\bibfnamefont {H.}~\bibnamefont {Kamada}}, \bibinfo
  {author} {\bibfnamefont {H.}~\bibnamefont {Witala}}, \ and\ \bibinfo {author}
  {\bibfnamefont {W.}~\bibnamefont {Gloeckle}},\ }\href {\doibase
  10.1103/PhysRevC.63.034006} {\bibfield  {journal} {\bibinfo  {journal} {Phys.
  Rev.}\ }\textbf {\bibinfo {volume} {C63}},\ \bibinfo {pages} {034006}
  (\bibinfo {year} {2001})},\ \Eprint {http://arxiv.org/abs/nucl-th/0008008}
  {arXiv:nucl-th/0008008 [nucl-th]} \BibitemShut {NoStop}%
\bibitem [{\citenamefont {De~Forest}(1983)}]{DeForest:1983vc}%
  \BibitemOpen
  \bibfield  {author} {\bibinfo {author} {\bibfnamefont {T.}~\bibnamefont
  {De~Forest}},\ }\href {\doibase 10.1016/0375-9474(83)90124-0} {\bibfield
  {journal} {\bibinfo  {journal} {Nucl. Phys.}\ }\textbf {\bibinfo {volume}
  {A392}},\ \bibinfo {pages} {232} (\bibinfo {year} {1983})}\BibitemShut
  {NoStop}%
\bibitem [{\citenamefont {Caballero}\ \emph {et~al.}(1993)\citenamefont
  {Caballero}, \citenamefont {Donnelly},\ and\ \citenamefont
  {Poulis}}]{Caballero:1992tt}%
  \BibitemOpen
  \bibfield  {author} {\bibinfo {author} {\bibfnamefont {J.~A.}\ \bibnamefont
  {Caballero}}, \bibinfo {author} {\bibfnamefont {T.~W.}\ \bibnamefont
  {Donnelly}}, \ and\ \bibinfo {author} {\bibfnamefont {G.~I.}\ \bibnamefont
  {Poulis}},\ }\href {\doibase 10.1016/0375-9474(93)90503-P} {\bibfield
  {journal} {\bibinfo  {journal} {Nucl. Phys.}\ }\textbf {\bibinfo {volume}
  {A555}},\ \bibinfo {pages} {709} (\bibinfo {year} {1993})}\BibitemShut
  {NoStop}%
\bibitem [{\citenamefont {Lachniet}\ \emph {et~al.}(2009)\citenamefont
  {Lachniet} \emph {et~al.}}]{Lachniet:2008qf}%
  \BibitemOpen
  \bibfield  {author} {\bibinfo {author} {\bibfnamefont {J.}~\bibnamefont
  {Lachniet}} \emph {et~al.} (\bibinfo {collaboration} {CLAS}),\ }\href
  {\doibase 10.1103/PhysRevLett.102.192001} {\bibfield  {journal} {\bibinfo
  {journal} {Phys. Rev. Lett.}\ }\textbf {\bibinfo {volume} {102}},\ \bibinfo
  {pages} {192001} (\bibinfo {year} {2009})},\ \Eprint
  {http://arxiv.org/abs/0811.1716} {arXiv:0811.1716 [nucl-ex]} \BibitemShut
  {NoStop}%
\bibitem [{\citenamefont {Venkat}\ \emph {et~al.}(2011)\citenamefont {Venkat},
  \citenamefont {Arrington}, \citenamefont {Miller},\ and\ \citenamefont
  {Zhan}}]{Venkat:2010by}%
  \BibitemOpen
  \bibfield  {author} {\bibinfo {author} {\bibfnamefont {S.}~\bibnamefont
  {Venkat}}, \bibinfo {author} {\bibfnamefont {J.}~\bibnamefont {Arrington}},
  \bibinfo {author} {\bibfnamefont {G.~A.}\ \bibnamefont {Miller}}, \ and\
  \bibinfo {author} {\bibfnamefont {X.}~\bibnamefont {Zhan}},\ }\href {\doibase
  10.1103/PhysRevC.83.015203} {\bibfield  {journal} {\bibinfo  {journal} {Phys.
  Rev.}\ }\textbf {\bibinfo {volume} {C83}},\ \bibinfo {pages} {015203}
  (\bibinfo {year} {2011})},\ \Eprint {http://arxiv.org/abs/1010.3629}
  {arXiv:1010.3629 [nucl-th]} \BibitemShut {NoStop}%
\bibitem [{\citenamefont {Arrington}(2012)}]{private_arrington}%
  \BibitemOpen
  \bibfield  {author} {\bibinfo {author} {\bibfnamefont {J.}~\bibnamefont
  {Arrington}},\ }\href@noop {} {\bibfield  {journal} {\bibinfo  {journal}
  {private communication}\ } (\bibinfo {year} {2012})}\BibitemShut {NoStop}%
\bibitem [{\citenamefont {Arrington}\ \emph {et~al.}(2007)\citenamefont
  {Arrington}, \citenamefont {Melnitchouk},\ and\ \citenamefont
  {Tjon}}]{Arrington:2007ux}%
  \BibitemOpen
  \bibfield  {author} {\bibinfo {author} {\bibfnamefont {J.}~\bibnamefont
  {Arrington}}, \bibinfo {author} {\bibfnamefont {W.}~\bibnamefont
  {Melnitchouk}}, \ and\ \bibinfo {author} {\bibfnamefont {J.~A.}\ \bibnamefont
  {Tjon}},\ }\href {\doibase 10.1103/PhysRevC.76.035205} {\bibfield  {journal}
  {\bibinfo  {journal} {Phys. Rev.}\ }\textbf {\bibinfo {volume} {C76}},\
  \bibinfo {pages} {035205} (\bibinfo {year} {2007})},\ \Eprint
  {http://arxiv.org/abs/0707.1861} {arXiv:0707.1861 [nucl-ex]} \BibitemShut
  {NoStop}%
\bibitem [{\citenamefont {Kelly}(2004)}]{Kelly:2004hm}%
  \BibitemOpen
  \bibfield  {author} {\bibinfo {author} {\bibfnamefont {J.~J.}\ \bibnamefont
  {Kelly}},\ }\href {\doibase 10.1103/PhysRevC.70.068202} {\bibfield  {journal}
  {\bibinfo  {journal} {Phys. Rev.}\ }\textbf {\bibinfo {volume} {C70}},\
  \bibinfo {pages} {068202} (\bibinfo {year} {2004})}\BibitemShut {NoStop}%
\bibitem [{\citenamefont {Jones}\ \emph {et~al.}(2000)\citenamefont {Jones}
  \emph {et~al.}}]{Jones:1999rz}%
  \BibitemOpen
  \bibfield  {author} {\bibinfo {author} {\bibfnamefont {M.~K.}\ \bibnamefont
  {Jones}} \emph {et~al.} (\bibinfo {collaboration} {Jefferson Lab Hall A}),\
  }\href {\doibase 10.1103/PhysRevLett.84.1398} {\bibfield  {journal} {\bibinfo
   {journal} {Phys. Rev. Lett.}\ }\textbf {\bibinfo {volume} {84}},\ \bibinfo
  {pages} {1398} (\bibinfo {year} {2000})},\ \Eprint
  {http://arxiv.org/abs/nucl-ex/9910005} {arXiv:nucl-ex/9910005 [nucl-ex]}
  \BibitemShut {NoStop}%
\end{thebibliography}%

\end{document}